\documentclass[twocolumn, times]{aastex631}

\usepackage{booktabs}
\usepackage{multirow}
\usepackage{amsmath}
\begin{document}
\hypersetup{colorlinks=true, citecolor=blue, urlcolor=blue, linkcolor=blue, allcolors = blue}

\title{Machine Learning Classification of COSMOS2020 Galaxies: Quiescent vs. Star-Forming}

\author{Vahid Asadi~\href{https://orcid.org/0009-0005-8897-2385}{\includegraphics[scale=0.04]{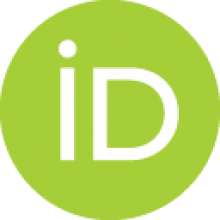}}}
\affiliation{Department of Physics, Institute for Advanced Studies in Basic Sciences (IASBS), PO Box 11365-9161, Zanjan, Iran; \url{vahidasadi@iasbs.ac.ir}}

\author{Nima Chartab~\href{https://orcid.org/0000-0003-3691-937X}{\includegraphics[scale=0.04]{orcid-ID.png}}}
\affiliation{Infrared Processing and Analysis Center, California Institute of Technology, Pasadena, CA 91125, USA}

\author{Akram Hasani Zonoozi~\href{https://orcid.org/0000-0002-0322-9957}{\includegraphics[scale=0.04]{orcid-ID.png}}}
\affiliation{Department of Physics, Institute for Advanced Studies in Basic Sciences (IASBS), PO Box 11365-9161, Zanjan, Iran; \url{vahidasadi@iasbs.ac.ir}}
\affiliation{Helmholtz-Institut f\"ur Strahlen-und Kernphysik (HISKP), Universit\"at Bonn, Nussallee 14-16, D-53115 Bonn, Germany; \url{haghi@iasbs.ac.ir}}

\author{Hosein Haghi~\href{https://orcid.org/0000-0002-9058-9677}{\includegraphics[scale=0.04]{orcid-ID.png}}}
\affiliation{Department of Physics, Institute for Advanced Studies in Basic Sciences (IASBS), PO Box 11365-9161, Zanjan, Iran; \url{vahidasadi@iasbs.ac.ir}}
\affiliation{Helmholtz-Institut f\"ur Strahlen-und Kernphysik (HISKP), Universit\"at Bonn, Nussallee 14-16, D-53115 Bonn, Germany; \url{haghi@iasbs.ac.ir}}
\affiliation{School of Astronomy, Institute for Research in Fundamental Sciences (IPM), PO Box 19395 - 5531, Tehran, Iran}

\author{Ghassem Gozaliasl~\href{https://orcid.org/0000-0002-0236-919X}{\includegraphics[scale=0.04]{orcid-ID.png}}}
\affiliation{Department of Computer Science, Aalto University, PO Box
15400, Espoo 00 076, Finland}
\affiliation{Department of Physics, University of Helsinki, PO Box 64,
00014 Helsinki, Finland}

\author{Aryana Haghjoo}
\affiliation{Department of Physics and Astronomy, University of California, Riverside, Riverside, CA 92521, USA}

\author{Bahram Mobasher~\href{https://orcid.org/0000-0001-5846-4404}{\includegraphics[scale=0.04]{orcid-ID.png}}}
\affiliation{Department of Physics and Astronomy, University of California, Riverside, Riverside, CA 92521, USA}

\begin{abstract}

Accurately distinguishing between quiescent and star-forming galaxies is essential for understanding galaxy evolution. Traditional methods, such as spectral energy distribution (SED) fitting, can be computationally expensive and may struggle to capture complex galaxy properties. This study aims to develop a robust and efficient machine learning (ML) classification method to identify quiescent and star-forming galaxies within the \texttt{Farmer} COSMOS2020 catalog. We utilized JWST wide-field light cones from the Santa Cruz semi-analytical modeling framework to train a supervised ML model, the \texttt{CatBoostClassifier}, using 28 color features derived from 8 mutual photometric bands within the COSMOS catalog. The model was validated against a testing set and compared to the SED-fitting method in terms of precision, recall, F1-score, and execution time. Preprocessing steps included addressing missing data, injecting observational noise, and applying a magnitude cut ($m_{\text{ch1}} < 26$ AB) along with a redshift range of $0.2 < z < 3.5$ to align the simulated and observational datasets. The ML method achieved an F1-score of 89\% for quiescent galaxies, significantly outperforming the SED-fitting method, which achieved 54\%. The ML model demonstrated superior recall (88\% vs. 38\%) while maintaining comparable precision. When applied to the COSMOS2020 catalog, the ML model predicted a systematically higher fraction of quiescent galaxies across all redshift bins within $0.2<z<3.5$ compared to traditional methods like NUVrJ and SED-fitting. This study shows that ML, combined with multi-wavelength data, can effectively identify quiescent and star-forming galaxies, providing valuable insights into galaxy evolution. The trained classifier and full classification catalog are publicly available.

\end{abstract}


\section{Introduction}
The classification of quiescent galaxies is crucial for understanding galaxy evolution. These galaxies represent a phase in which star formation has ceased, providing a natural laboratory to study the processes that lead to the quenching of star formation. Over the last decade, the existence of this population of galaxies at high redshift ($z\gtrsim2$) has been firmly established \citep[e.g.,] []{glazebrook2017massive, schreiber2018near, girelli2019massive, santini2021emergence, carnall2023surprising, nanayakkara2024population, kakimoto2024massive}. Quiescent galaxies were initially identified by their redder colors and Balmer break in their spectra, which indicate the presence of older stellar populations with little or no ongoing star formation. In contrast to star-forming galaxies, which typically exhibit emission lines from young, hot stars, the stellar populations in quiescent galaxies are predominantly composed of old, evolved stars.

Classifying galaxies into quiescent and star-forming categories provides valuable insights into the processes that shape the life cycle of galaxies and the universe as a whole. The mechanisms behind the quiescence of galaxies remain a topic of ongoing debate. It is believed that quiescence can arise from various processes, including galaxy mergers that strip a galaxy of its gas, thereby halting star formation \citep{Hopkins2008}; feedback from supermassive black holes, which can regulate star formation by expelling gas from the galaxy \citep{Fabian2012, King2015}; and environmental effects, such as those observed in galaxy clusters, where gas may be stripped away from galaxies in high-density environments \citep{Boselli2006}.

The cosmic star formation rate appears to peak at a redshift of $z\sim2$ and subsequently declines toward the present day \citep{Madau2014}. Understanding the properties of quenched and star-forming galaxies across different redshifts reveals how star formation activity evolves throughout the universe. The fraction of quenched galaxies, along with the time evolution of this ratio, provides important constraints for galaxy formation models and cosmological simulations \citep{Peng2010}.

The identification of quiescent galaxies presents a significant challenge, primarily due to inherent degeneracies among physical parameters such as stellar age, metallicity, and dust reddening \citep{worthey1994comprehensive}. These degeneracies can result in the misclassification of galaxies, as their broadband spectral energy distributions (SEDs) often overlap due to similar observed characteristics. Furthermore, the absence of redshift information exacerbates these classification issues by hindering the determination of distances and cosmic evolution timelines.

The UVJ color-color selection technique is a widely used method for identifying quiescent galaxies \citep[e.g.,] []{strateva2001color,williams2009detection,van20143d, shahidi2020selection}. Other alternative color combinations, such as those used by \cite{daddi2004new} and \cite{leja2019beyond}, have also been explored. However, these methods, including UVJ, simplify the data by focusing on a few color indices, potentially neglecting valuable information contained within the broader SED. To compensate for this potential loss of accuracy, galaxies identified as quiescent candidates through color selection are often further analyzed using SED-fitting \citep[e.g.,] []{wiklind2008population,girelli2019massive} and spectroscopic observations \citep[e.g.,] []{glazebrook2017massive,schreiber2018near}. While these techniques provide deeper insights into galaxy properties and confirm quiescent status, they can be time-consuming and resource-intensive, especially for large-scale surveys.  

In contrast, machine learning (ML) techniques offer a more promising approach for selecting quiescent galaxies. By leveraging the full richness of multi-waveband data, these methods can learn complex patterns and relationships within the SED that are not easily captured by simple color-color criteria. This capability enables them to make more accurate and precise selections, thereby improving both completeness and purity \citep{steinhardt2020method,humphrey2023euclid}.

In this study, we utilize a supervised ML model to classify quiescent galaxies from star-forming ones within the \texttt{Farmer} COSMOS2020 catalog \citep{weaver2022cosmos2020}. This catalog is ideal for our purposes as it combines deep, multi-wavelength photometry and a large contiguous field, which are critical for robust population statistics and SED-based classification.

The model was trained on colors constructed from selected band sets from the COSMOS field in the Santa Cruz semi-analytic James Webb Space Telescope \citep[JWST; ][]{gardner2006james} wide-field light cones \citep{somerville2021mock,yung2022semi}. These mock observations provide physically realistic galaxy properties that are essential for supervised learning classification. To enhance the model's robustness, observational noise was injected into the training data.

\begin{figure*}
    \centering
    \includegraphics[width=0.9\linewidth]{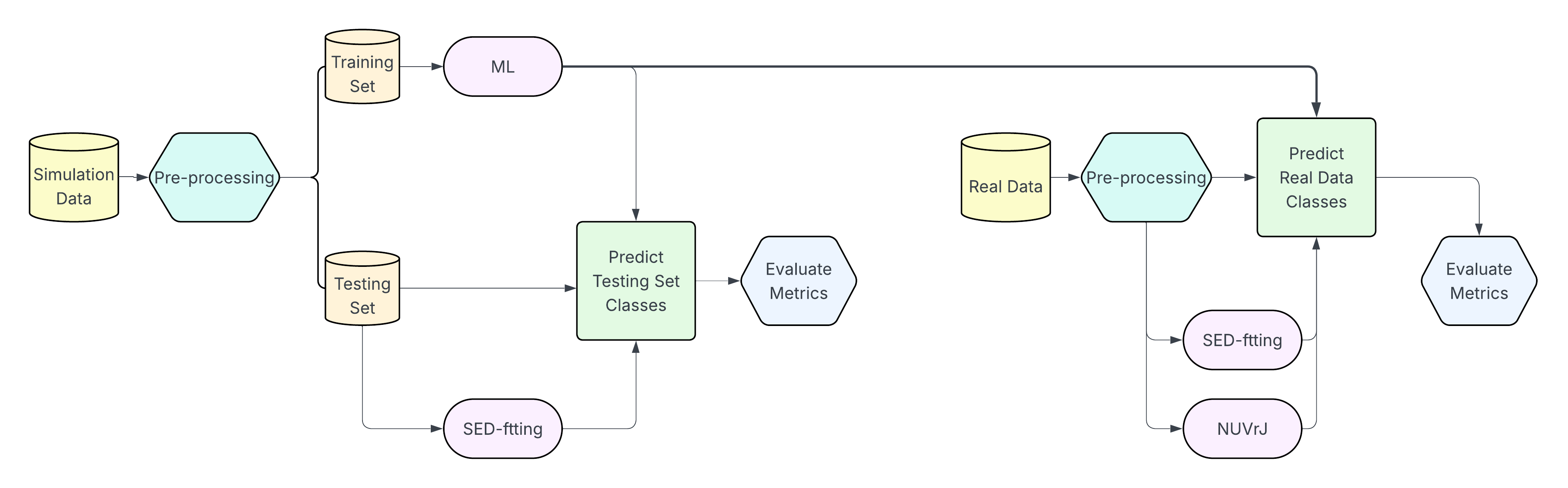}
    \caption{Overview of the data analysis pipeline used in this study.}
    \label{fig:fig_flow}
\end{figure*}

The paper is structured as follows: Section~\ref{sec:2} introduces the mock photometry catalog and the COSMOS2020 catalog. Figure~\ref{fig:fig_flow} illustrates our end-to-end analysis pipeline, which integrates mock simulation and observational data via ML classification. The methodology is detailed progressively in subsequent sections. Section~\ref{sec:3} details the pre-processing steps applied to both observational and simulated data. Section~\ref{sec:4} outlines the ML and SED-fitting methods used for classifying galaxies as quiescent or star-forming. Section~\ref{sec:5} compares the performance of these two classification approaches. Section~\ref{sec:6} discusses the potential reasons for the superiority of the ML method over the SED-fitting method, identify potential limitations of both approaches, and suggests avenues for future improvement. In Section~\ref{sec:7} we apply our ML model on the COSMOS2020 catalog and classify its galaxies into quiescent and star-forming categories, and compare the results with those from the NUVrJ and SED-fitting methods. Finally, Section~\ref{sec:8} provides an overview of our study and draws conclusions.

We adopt a flat $\Lambda$CDM cosmology with parameters $H_{0}=70kms^{-1}Mpc^{-1}$, $\Omega_{m}=0.3$ and $\Omega_{\Lambda}=0.7$. All magnitudes are reported in the AB system \citep{oke1983secondary}.


\section{Data}\label{sec:2}

\subsection{Mock galaxy catalog}
Mock galaxy catalogs are synthetic datasets designed to replicate the observable properties of real galaxies and their large-scale distribution in the universe. By integrating cosmological simulations with galaxy formation models, these catalogs create a controlled framework for testing hypotheses, comparing observations, and refining theoretical models. This approach effectively bridges the gap between theoretical predictions and observational data.

For this study, we utilized JWST wide-field light cones, which are based on an updated version of the Santa Cruz semi-analytic modeling (SAM) framework \citep{somerville2021mock,yung2022semi}. The SAM provides a computational framework for studying galaxy formation and evolution by modeling the complex physical processes that govern galaxies within dark matter halos. In this framework, dark matter merger trees are traced back to their minimum progenitor mass using either cosmological N-body simulations or analytical approaches, such as the Extended Press-Schechter (EPS) formalism \citep{press1974formation}. We employed the high-resolution Bolshoi-Planck cosmological simulation \citep{klypin2011dark}, which evolves dark matter particles under the $\Lambda$CDM cosmological model, offering a detailed view of galaxy formation over time. By combining these merger trees with parameterized prescriptions for key astrophysical processes—such as gas accretion, cooling, star formation, feedback, and galaxy mergers—the SAM predicts a wide range of galaxy properties, including stellar mass, star formation rates, metallicity, and luminosity. This allows for a comprehensive examination of galaxy formation over time.

To mimic the observational conditions of deep-field surveys, the dataset is structured into light cones that represent the three-dimensional spatial distribution of galaxies across cosmic time. These light cones include five independent fields with coordinates and geometries overlapping the five Cosmic Assembly Near-IR Deep Extragalactic Legacy Survey \citep[CANDELS; ][]{grogin2011candels,koekemoer2011candels} legacy fields (GOODS-S, GOODS-N, COSMOS, EGS, and UDS). Each field has eight realizations, covering redshift values from z=0 to z=10. This catalog provides observed-frame photometry for a wide range of telescopes, including JWST (NIRCam) \citep{gardner2006james}, Roman (WFI) \citep{spergel2015wide}, Hubble (WFC3/ACS) \citep{dressel2012wide}, Spitzer \citep{werner2004spitzer}, Euclid \citep{laureijs2011euclid}, Rubin \citep{ivezic2019lsst}, GALEX \citep{martin2005galaxy}, SDSS \citep{york2000sloan}, UKIRT \citep{lawrence2007ukirt}, VISTA \citep{emerson2006visible}, and DECam \citep{flaugher2015dark}. Additionally, it offers rest-frame luminosities for NUV, FUV, and Johnson/Bessel/Cousins bands, along with various predicted physical properties (e.g., mass, star formation rate, metallicity) for halos and galaxies.

In this framework, galaxies are populated within dark matter halos, with the most massive galaxy classified as the central galaxy, while the others are referred to as satellite galaxies. Satellite galaxies experience significant mass loss due to tidal stripping and dynamical friction, which can lead to their destruction before they merge with the central galaxy \citep{somerville2008semi}. The gas accreted onto these halos is shock-heated and subsequently cools, with the cooling process modeled based on a spherically symmetric density profile \cite{white1991galaxy}. Gas accretion is categorized into hot and cold modes, both of which influence star formation processes. The rate of gas cooling is regulated by the ratio of the cooling time to the free-fall time. Only gas that cools efficiently within a halo's dynamical timescale contributes to the star-forming reservoir.

Star formation occurs through two primary modes: quiescent star formation, which utilizes an updated Kennicutt-Schmidt relation focusing on $H_{2}$ surface density \citep{somerville2015star}, and merger-driven starbursts, where rapid gas compression enhances star formation. The latter is modeled based on binary galaxy merger simulations \citep[i.e.] []{robertson2006fundamental, hopkins2009disks}. The model also incorporates feedback mechanisms from massive stars and supernovae, which can expel cold gas from galaxies, thereby affecting their evolution \citep{somerville2008semi}.

Additionally, each galaxy is assumed to host a seed central black hole that can grow through two modes: a quasar mode, driven by mergers and disk instabilities, and a radio mode, involving the accretion of hot halo gas, following the Bondi-Hoyle-Lyttleton model \citep{bondi1952spherically}. 
The parameters governing these processes are calibrated against observational data, such as galaxy mass functions, luminosity functions, and scaling relations (e.g., the Tully-Fisher relation or the black hole–bulge mass relation). This calibration ensures that the SAM-generated spectral energy distributions (SEDs) align with real observations. These SEDs are further adjusted for dust attenuation and intergalactic medium absorption \citep{madau1995radiative} and are convolved with filter response functions to derive fluxes and magnitudes in each observational band \citep[e.g.,][]{yung2019semi, somerville2021mock, yung2022semi}.

\subsection{Observational galaxy catalog}
The COSMOS survey covers an area of 2 $deg^{2}$ and provides a wealth of deep multiwavelength data across over 40 bands from major global facilities \citep{scoville2007cosmic}. The COSMOS2020 source catalog \citep{weaver2022cosmos2020} represents a significant advancement over its predecessor \citep{laigle2016cosmos2015}, offering detections along with new ultra-deep optical/NIR imaging and multiwavelength photometry for 1.7 million sources throughout the entire COSMOS field. Notably, approximately 89,000 of these sources have measurements in all available broadband filters.

Key enhancements in COSMOS2020 include new ultra-deep optical data from the Hyper Suprime-Cam Subaru Strategic Program (HSC-SSP) Public Data Release 2 (PDR2) \citep{aihara2019second}, fresh data from the Visible Infrared Survey Telescope for Astronomy (VISTA) Data Release 4 (DR4) \citep{mccracken2012ultravista}, which extends coverage one magnitude deeper over the entire area compared to previous datasets, and the inclusion of all Spitzer IRAC data in COSMOS \citep{moneti2022euclid}. Additionally, legacy datasets, such as Suprime-Cam imaging, have been reprocessed (see \cite{weaver2022cosmos2020} for details).

COSMOS2020 comprises two independent photometric catalogs. The first is the \texttt{Classic} catalog, which employs standard aperture photometry \citep{bertin1996sextractor} on point-spread function (PSF)-homogenized optical/NIR images. For IRAC images, the \texttt{IRACLEAN} software \citep{hsieh2012taiwan} is utilized for PSF photometry. The second catalog is generated using The \texttt{Farmer}, a novel profile-fitting photometric software developed specifically for COSMOS2020. By leveraging The \texttt{Tractor} software \citep{lang2016tractor} for source modeling, The \texttt{Farmer} produces reproducible source detection and photometry, resulting in a comprehensive multiwavelength catalog.

In this paper, we utilize The \texttt{Farmer} catalog throughout, as it provides more accurate photometry in various bands for fainter sources compared to the \texttt{Classic} catalog.


\begin{table}[t]
\centering
\caption{Overview of the 8 mutual bands between the SAM and COSMOS2020 catalogs, spanning from UV to NIR.}
\setlength{\belowcaptionskip}{10pt}
\begin{tabular}{ccccc}
\hline \hline
\toprule

Instrument                & Filter  & Central                 & Width     & $3\sigma$ depth \\
/Telescope                &         &$\lambda$[{\AA}]          & [{\AA}]   & ($3^{\prime\prime}/2^{\prime\prime}$) \\
(Survey)                  &         &                         &           & $\pm$ 1         \\
\midrule
MegaCam/CFHT       & u        & 3709                  & 518       & 27.8/27.2                \\
\midrule
ACS/HST            & F814W    & 8333                  & 2511      & 27.8                     \\
\midrule
VIRCAM             & Y        & 10216                 & 923       & 25.3/24.8              \\
/UltraVISTA        & J        & 12525                 & 1718      & 25.2/24.7              \\
                   & H        & 16466                 & 2905      & 24.9/24.4              \\
                   & $K_{s}$  & 21557                 & 3074      & 25.3/24.8              \\
\midrule
IRAC/\textit{Spitzer}     & ch1     & 35686             & 7443      & 26.4/25.7               \\
                          & ch2     & 45067             & 10119     & 26.3/25.6               \\
\bottomrule
\label{tab:tab0}
\end{tabular}
\end{table}

\begin{figure}
    \centering
    \includegraphics[width=0.95\linewidth]{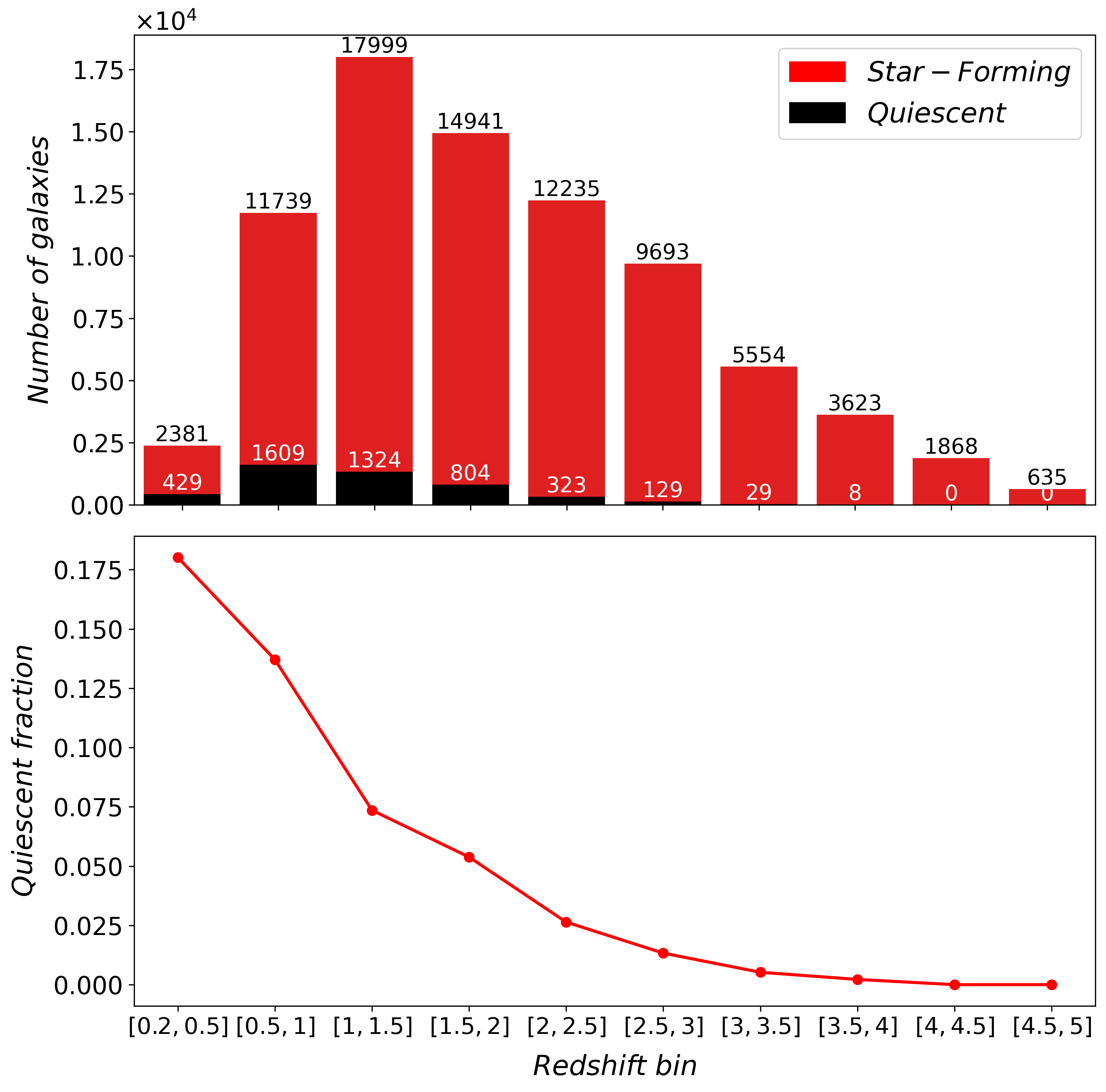}
    \caption{Distribution of total galaxies and the quiescent galaxy fraction as a function of redshift bins in the mock sample for $M_{QG} \geq 10^{9.5} M_{\odot}$. The plot highlights the significant decline in the fraction of quiescent galaxies at  $z > 3.5$.}
    \label{fig:fig1}
\end{figure}

\section{Pre-processing}\label{sec:3}
In this section, we outline the steps we took to prepare both the mock and observational data before any operations were performed.

\subsection{Selecting mutual bands}
We identified 8 mutual bands between the SAM catalog and the COSMOS2020 catalog. These bands include one CFHT filter (u), one ACS filter (F814W), four UVISTA filters (Y, J, H, and $K_{s}$), and two $Spitzer$/IRAC filters (ch1 and ch2). Table~\ref{tab:tab0} shows these bands in details.

\subsection{Sample selection}
For observational data, we utilize the galaxy sample originally constructed by \cite{weaver2023cosmos2020}. This sample selected from the contiguous 1.27 deg$^2$ COMBINED region, which is defined as the union of the UltraVISTA and Subaru/SC footprints, excluding regions around bright stars. To ensure robust measurements of stellar masses and redshifts, \cite{weaver2023cosmos2020} applied several quality cuts, including the removal of sources with low infrared detection ($m_{\text{ch1}} > 26$ AB), ambiguous redshifts ($> 32\%$ of redshift probability outside $z_{\text{phot}} \pm 0.5$), and unreliable SED fits (reduced $\chi^2 > 10$).

For the mock catalog, we focused specifically on the COSMOS field and its its zeroth realization. Similar to the observational data, we excluded sources with low infrared detection ($m_{\text{ch1}} > 26$ AB).

Our analysis focused on galaxies within the redshift range $0.2 < z < 3.5$ for both sample data. This restriction avoids the issue of very few quiescent galaxies at $z > 3.5$, which would negatively impact our ML method for quiescent galaxy classification. We will discuss this further later (Section~\ref{sub:sub1}).

After implementing these sample selection steps, we obtained 114,995 samples in the mock galaxy catalog and 365,877 samples in the observational galaxy catalog.

\begin{table}[t]
\centering
\caption{Percentage of missing values for magnitudes and fluxes in the COSMOS2020 sample across all used bands.}
\setlength{\belowcaptionskip}{10pt}

\begin{tabular}{c|c|c}
\hline
\hline
\toprule
COSMOS2020 bands used & \multicolumn{2}{c}{Missing value percentage} \\
\cmidrule{2-3}

                      & Magnitude & Flux \\
\hline
u        & 13\%       & 06\% \\
F814W    & 16\%       & 16\% \\
Y        & 03\%       & 00\% \\
J        & 02\%       & 00\% \\
H        & 02\%       & 00\% \\
$K_{s}$  & 01\%       & 00\% \\
ch1      & 00\%       & 00\% \\
ch2      & 02\%       & 01\% \\

\bottomrule
\end{tabular}
\label{tab:tab00}
\end{table}

\subsection{Mass completeness limits}\label{sub:sub0}
To ensure the completeness of our galaxy samples, we adopt the mass completeness limits for star-forming and quiescent galaxies derived by \cite{weaver2023cosmos2020}. These limits account for the varying detectability of different galaxy populations across redshift and are valid for the redshift range \(0.2 < z \leq 5.5\). The mass completeness limits are given by the following equations:

\begin{itemize}
    \item Star-Forming Galaxies:
    \begin{equation}
        M_{\text{lim}} = -5.77 \times 10^7 (1 + z) + 8.66 \times 10^7 (1 + z)^2
        \label{eq0}
    \end{equation}
    
    \item Quiescent Galaxies:
    \begin{equation}
        M_{\text{lim}} = -3.79 \times 10^8 (1 + z) + 2.98 \times 10^8 (1 + z)^2
        \label{eq00}
    \end{equation}
\end{itemize}

\subsection{Setting labels}\label{sub:sub1}
To identify the quiescent population at different epochs in the mock catalog, we adopted the evolving specific star-formation rate (sSFR) threshold criterion introduced by \cite{pacifici2016evolution}. The threshold is defined as follows:

\begin{equation}
    sSFR \leq sSFR_{lim} = \frac{0.2}{t_{U}(z)}
\label{eq1}
\end{equation}

In this equation, $sSFR$ represents the specific star-formation rate (in $Gyr^{1}$), and $t_{U}(z)$ denotes the age of the universe at a given redshift $z$ (in $Gyr$). This approach allows for a dynamic threshold that evolves with cosmic time, reflecting the declining star formation activity in galaxies as the universe ages.

By applying this criterion alongside the mass completeness limits (Equations~\ref{eq0} and \ref{eq00}) to galaxies within the redshift range $0.2 < z < 3.5$, we identified 7,034 quiescent galaxies and 69,894 star-forming galaxies. This resulted in an imbalanced dataset. As detailed in Section \ref{sec:5.1}, we address this imbalance using metrics such as the F1-score (Equation \ref{eq:3}). We limited our analysis to this redshift range because, as illustrated in Figure~\ref{fig:fig1}, the fraction of quiescent galaxies relative to the total galaxy population becomes negligible at $z > 3.5$.

\subsection{Filling missing values}
While the mock sample contained no missing values in the selected bands, the COSMOS2020 sample had missing data in all bands except for ch1 (as the sample is selected based on ch1). Table~\ref{tab:tab00} summarizes the percentage of missing values for both magnitudes and fluxes. To address the missing values in the remaining seven bands, we employed the \texttt{MissForest} method\footnote{\url{https://pypi.org/project/MissForest}} \citep{stekhoven2015missforest}. \texttt{MissForest} is a non-parametric imputation technique that utilizes Random Forests \citep{Breiman2001,biau2016random} to iteratively fill in missing entries.

\begin{figure}
    \centering
    \includegraphics[width=0.98\linewidth]{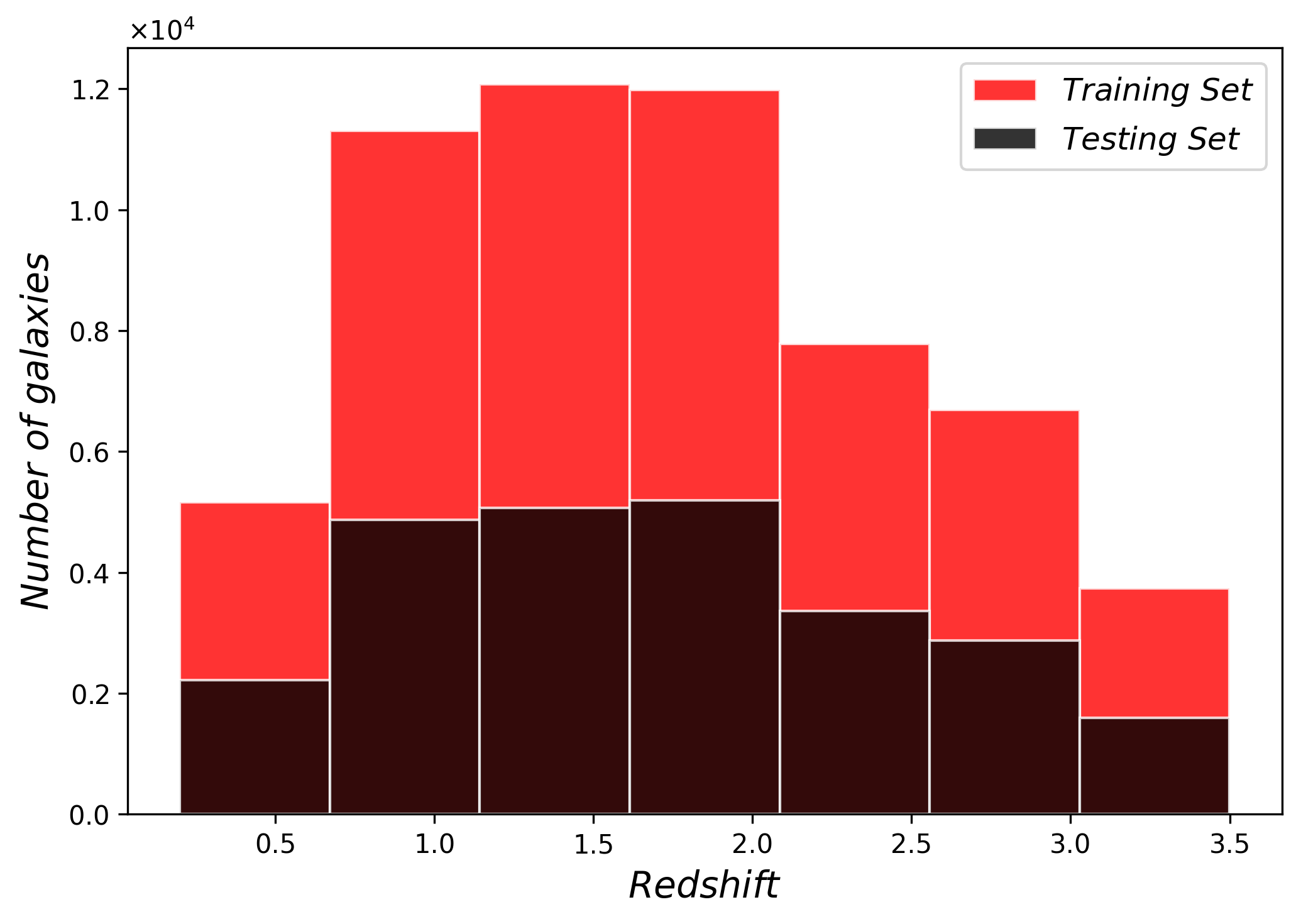}
    \caption{Redshift distributions of training and testing sets.}
    \label{fig:fig3_redshift}
\end{figure}

\begin{figure*}
    \centering
    \includegraphics[width=0.89\linewidth]{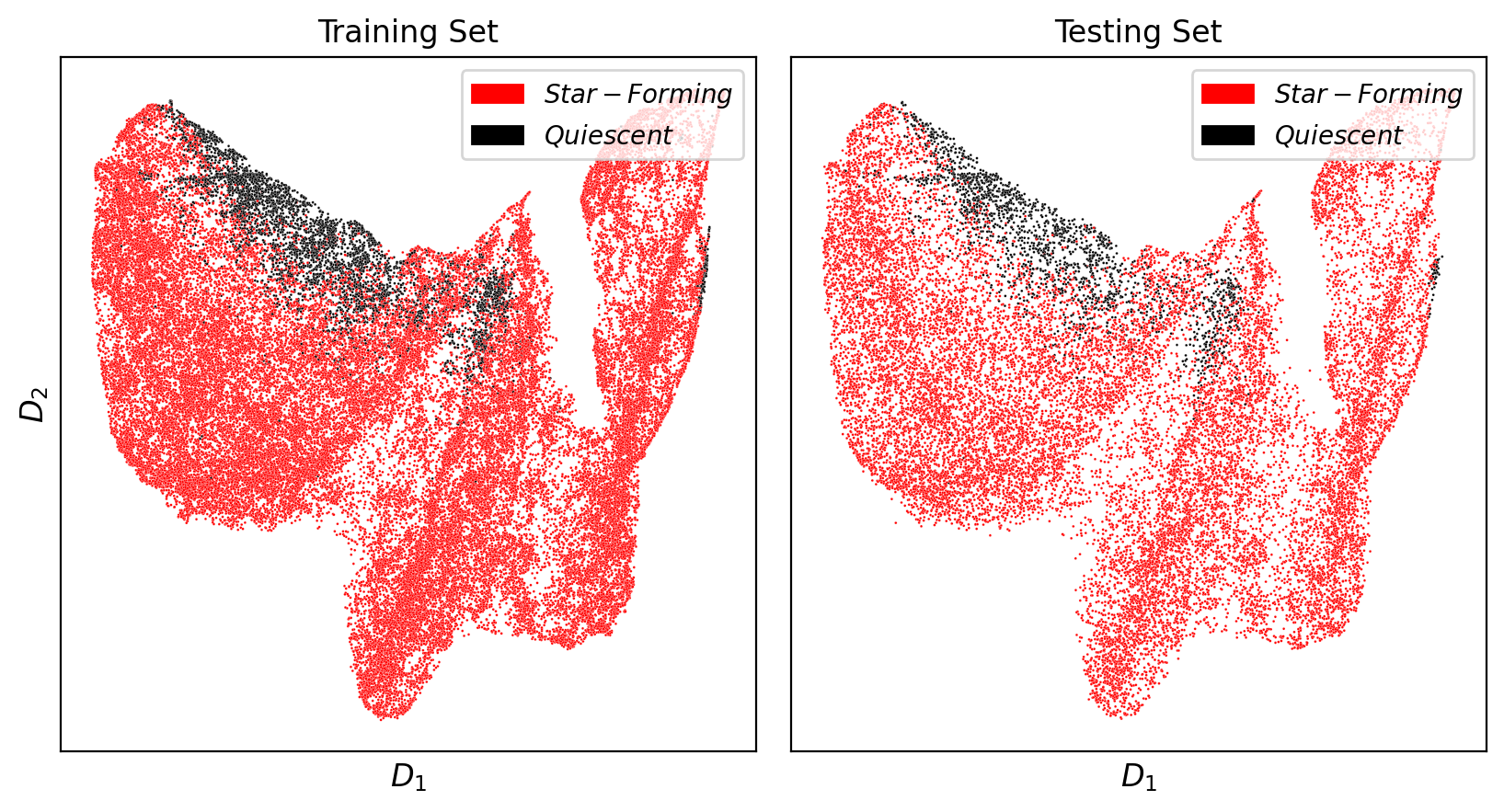}
    \caption{Distribution of training and testing sets in the mock sample, visualized using UMAP. Red and black points} represent star-forming and quiescent galaxies, respectively. The dimensions of the 2D plots (D1 and D2) are arbitrary labels and do not carry any physical significance.
    \label{fig:fig2}
\end{figure*}

We specifically used \texttt{MissForest} to fill in missing values for magnitudes, fluxes, and errors related to the fluxes. The accurate estimation of these values is crucial for subsequent analyses, as they impact the accuracy of our measurements and interpretations. During each iteration, a Random Forest model predicts the missing values for a specific column by leveraging the relationships between different variables in the dataset. These predictions are then incorporated back into the dataset. This iterative process continues until the model converges, ensuring that the imputed values are statistically robust and reflective of the underlying data distribution.

\subsection{Injecting noise}
To align the magnitude distributions of the mock sample with the COSMOS2020 data, we assigned uncertainties to each band's photometry based on its flux measurement, utilizing the uncertainties provided in the COSMOS2020 sample. For galaxies with negative fluxes in the observational data, we initially set these values to the limiting flux for the corresponding band.

Next, we trained a \texttt{RandomForestRegressor} model\footnote{\url{https://scikit-learn.org} \citep{pedregosa2011scikit}} \citep{Breiman2001,biau2016random} for each selected band using the COSMOS2020 data. In this model, we used magnitudes as features and the corresponding multiplicative flux errors as targets. This approach allowed us to capture the complex relationships between magnitudes and their associated uncertainties effectively.

The trained model was then used to predict multiplicative flux errors for the corresponding magnitude bands in the mock sample. To simulate observational errors, we drew a random number, denoted as $\delta F$, from a Gaussian distribution with a scale length defined by the predicted multiplicative uncertainty. We then added the term $\eta = -2.5 log_{10}(1 + \delta F)$ to account for these simulated observational errors, thereby enhancing the realism of our mock data \citep{simet2021comparison}.

\begin{figure*}
    \centering
    \includegraphics[width=\linewidth]{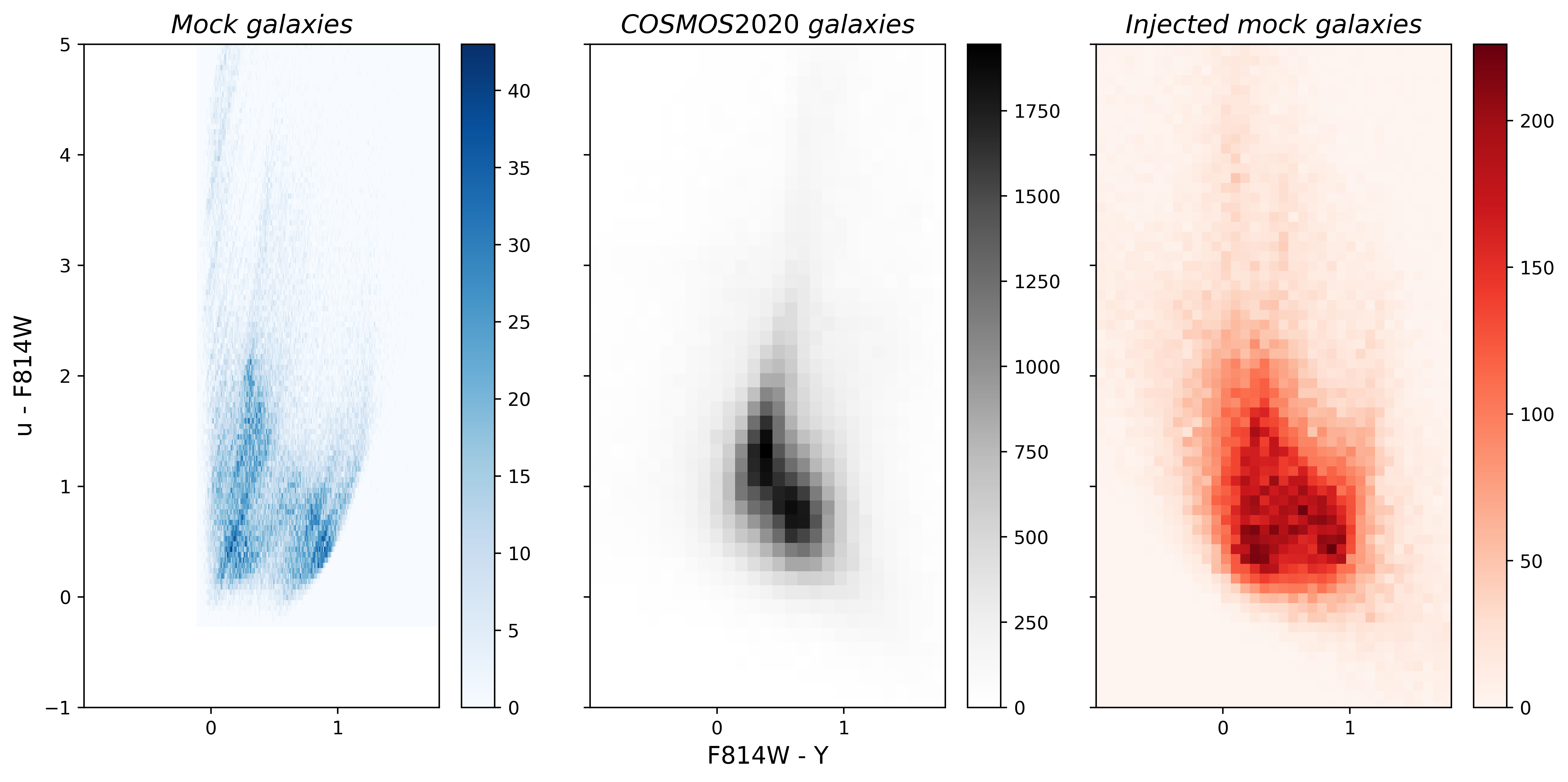}
    \caption{Heatmap comparison of the mock (with and without injected noise) and COSMOS2020 galaxy colors (u-F814W and F814W-Y). The injected mock galaxies show similar color distributions and cover the COSMOS2020 galaxies manifold, indicating good representation of real colors.}
    \label{fig:fig3}
\end{figure*}


\section{Classification Methods}\label{sec:4}
In this section, we explain our ML method and SED-fitting method for classifying quiescent and star-forming galaxies, and build a ground base for comparing these two approaches.
 
To achieve this goal, we first divided the injected mock data into two sets: a training set (70\%) and a testing set (30\%). The resulting training set had a size of 53,850, while the testing set contained 23,078 data points. To ensure the training and testing sets represent the same distribution of data, we stratified the data based on labels and redshifts. We used bins of width 0.5 to group the redshifts. Figure~\ref{fig:fig3_redshift} confirms the identical redshift distributions between both sets achieved through this stratification. Figure~\ref{fig:fig2} visually compares the training and testing sets using unsupervised Uniform Manifold Approximation and Projection\footnote{\url{https://pypi.org/project/umap-learn}} \citep[UMAP; ][]{mcinnes2018umap,healy2024uniform}.

It is important to note that the resulting UMAP dimensions (D1 and D2 for a 2D embedding) are nonlinear embeddings determined solely by the algorithm's optimization process to preserve local and global data structure. These dimensions carry no direct physical meaning. UMAP determines these dimensions by minimizing the cross-entropy between the high-dimensional color space topology and its low-dimensional representation.

\subsection{Machine Learning Method}
We utilized a ML algorithm known as the \texttt{CatBoostClassifier} algorithm\footnote{\url{https://catboost.ai}} \citep{prokhorenkova2018catboost,hancock2020catboost} for our classification approach. The CatBoost algorithm has been successfully applied to a variety of astronomical tasks. For example, it has been used to classify astronomical objects, such as distinguishing between quasars and galaxies \citep{hughes2022quasar}, and for identifying Fermi-LAT unidentified gamma-ray sources \citep{coronado2022classification}. The algorithm has also proven valuable in classifying quiescent and star-forming galaxies \citep{humphrey2023euclid} and has been deployed in other astronomical studies \citep[e.g., ][]{cunha2022photometric,coronado2023redshift,zeraatgari2024exploring,boulet2024catalogue,li2025application}.

\texttt{CatBoostClassifier} is a sophisticated algorithm designed for classification tasks, leveraging the power of gradient boosting to achieve high predictive accuracy. It constructs an ensemble of decision trees, where each successive tree is trained to correct the errors of its predecessors, thereby refining the model's predictions iteratively. This process involves incrementally minimizing the loss function, which enhances the model's ability to generalize from the training data. To understand this process more clearly, consider a golfer. The model's training is like a series of swings. Initially, a golfer might have a less-than-perfect swing. Each successive shot is a chance to learn from the last one's mistakes, making small adjustments to the grip, stance, and follow-through to improve accuracy \citep{harrison2023effective}.

A distinctive feature of the \texttt{CatBoostClassifier} is its implementation of ordered boosting. This is a novel and crucial technique that directly addresses the problem of overfitting, which occurs when a model learns the training data too well and fails to generalize to new, unseen data. By carefully balancing model complexity with the available data, CatBoost's ordered boosting ensures the model remains robust and effective, delivering high performance and stable predictions even on diverse datasets. It achieves this by introducing a more principled and less biased way of calculating gradients, which are the signals the algorithm uses to learn, ultimately resulting in a model that is less prone to memorizing noise in the data \citep[see Section 4.2 in ][ for more details]{prokhorenkova2018catboost}.

While the \texttt{RandomForestClassifier} and other popular ensemble methods like \texttt{XGBoostClassifier} \citep[e.g., ][]{bentejac2021comparative} also utilize an ensemble of trees, they differ in their approach. The \texttt{RandomForestClassifier} builds multiple independent trees on bootstrapped samples of the data and combines their predictions through a voting process. In contrast, both CatBoost and XGBoost are gradient boosting algorithms that build trees sequentially, with each new tree correcting the errors of the previous ones. However, CatBoost's key advantage lies in its novel approach to handling categorical features and its ordered boosting mechanism, which is specifically designed to more effectively prevent overfitting than standard gradient boosting methods like XGBoost.

For training our \texttt{CatBoostClassifier}, we constructed all the unique possible color combinations using the chosen bands (Table~\ref{tab:tab0}) in the injected mock catalog, resulting in 28 distinct colors. These include both adjacent bands (e.g., u - F814W, F814W - Y, Y - J) and widely separated bands (e.g., u - $K_{s}$, Y - ch1, F814W - ch2). 

Before training, we applied the \texttt{StandardScaler} function\footnote{\url{https://scikit-learn.org}}, which scales the data to have zero mean and unit variance:

\begin{equation}
X_{\text{scaled}} = \frac{X - \mu}{\sigma},
\end{equation}

where $\mu$ is the mean and $\sigma$ is the standard deviation of each color feature. This preprocessing step ensures that all color dimensions contribute equally during machine learning by eliminating arbitrary scale differences between them.

To validate that the constructed colors accurately represent the real colors in the observational catalog, we compared the heatmaps of two selected colors: u - F814W and F814W - Y for both mock galaxies (with and without simulated noise) and the observed galaxies, as shown in Figure~\ref{fig:fig3}. The results indicate that the injected mock galaxies occupy a similar region as the COSMOS galaxies and also cover their manifold, suggesting a good representation.

Optimization experiments did not yield significant performance improvements, so we opted for default settings. The trained model then labeled galaxies in the testing set as quiescent or star-forming.

\subsection{SED-fitting Method}
Using the selected bands, we employed SED-fitting code \texttt{LePhare}\footnote{\url{https://www.cfht.hawaii.edu/~arnouts/LEPHARE/lephare.html}} \citep{arnouts1999measuring,ilbert2006accurate} coupled with the \cite{bruzual2003stellar} stellar population synthesis models, to derive the median sSFR for each injected mock galaxy in the testing set. We assumed an exponentially declining star formation history (SFH) with nine different e-folding timescales ($\tau$) ranging from $0.01$ to $30$ $Gyr$. 

Dust properties are modeled with varying E(B-V) values between $0$ and $1$, considering the \cite{calzetti2000dust}. Nebular emission line contributions are included as described in \cite{ilbert2008cosmos}. A Chabrier initial mass function (IMF) \citep{chabrier2003galactic} is adopted with a mass range of $0.01$ to $100 M\odot$. Three different metallicity values are considered: $Z$ = $0.02$, $0.008$, and $0.004$. The redshift was fixed based on values taken from the mock catalog.

After obtaining the sSFR for each injected mock galaxy, we used Equation~\ref{eq1} to label the galaxies in the testing set.


\section{Results}\label{sec:5}

\begin{figure}
    \centering
    \includegraphics[width=\linewidth]{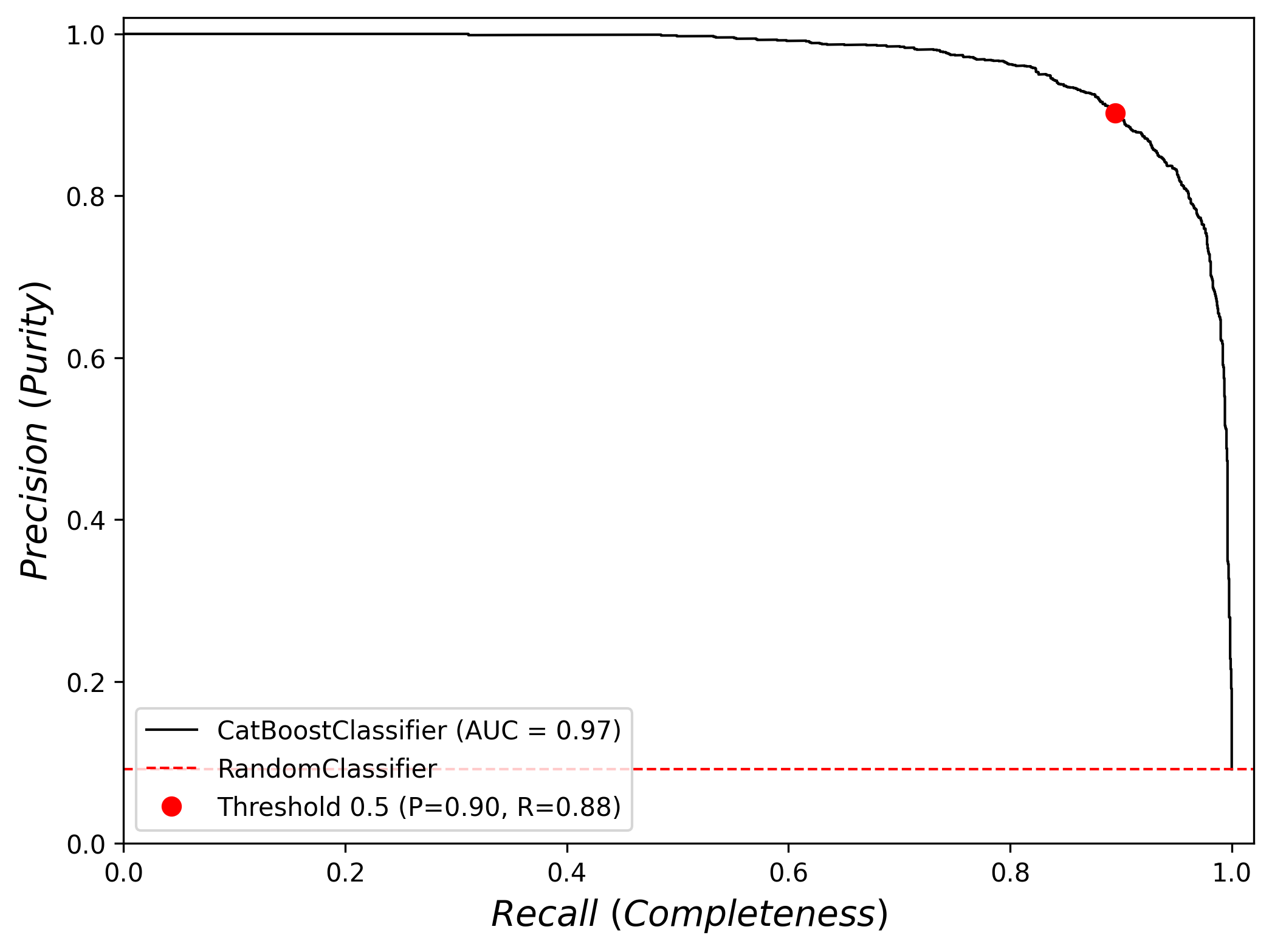}
    \caption{Precision-recall curve for quiescent galaxy classification (AUC = 0.97). The red circle marks our operating threshold (0.5), achieving 90\% precision and 88\% recall. The dashed line shows random-classifier performance.}
    \label{fig:fig4_roc}
\end{figure}

\subsection{Performance Metrics}\label{sec:5.1}
To evaluate and compare the performance of our classifiers, we utilized the confusion matrix \citep[e.g., see ] []{stehman1997selecting} along with derived metrics such as precision (purity), recall (completeness), and F1-score \citep[e.g., see ] []{powers2020evaluation}. These metrics provide a comprehensive understanding of how well our classifiers distinguish between quiescent and star-forming galaxies, especially in the context of our imbalanced dataset where accuracy can be misleading. We also adopted a default classification threshold of 0.5 for our binary classification task, as this provides a balanced trade-off between precision and recall.

The confusion matrix is a table used to describe the performance of a classification model on a set of test data for which the true values are known. The matrix itself includes four key terms:

\begin{itemize}
    \item True Positives (TP): The number of correctly predicted positive observations.
    \item True Negatives (TN): The number of correctly predicted negative observations.
    \item False Positives (FP): The number of incorrect positive predictions.
    \item False Negatives (FN): The number of incorrect negative predictions.
\end{itemize}

Table~\ref{tab:tab1} illustrates a confusion matrix for binary classification.

\begin{table}[h]
    \caption{Binary Confusion Matrix.}
    \centering
    \resizebox{0.45\textwidth}{!}{
    \begin{tabular}{c|c|c}
        & Predicted Positive & Predicted Negative \\
        \hline
        Actual Positive & TP & FN \\
        \hline
        Actual Negative & FP & TN \\
    \end{tabular}
    }
    \label{tab:tab1}
\end{table}

Precision is the ratio of correctly predicted positive observations to the total predicted positives. It is a measure of the accuracy of positive predictions.

\begin{equation}
\text{Precision} = \frac{TP}{TP + FP}
\label{eq:2}
\end{equation}

Recall is the ratio of correctly predicted positive observations to all actual positives. It shows how well the model can capture positive instances.

\begin{equation}
\text{Recall} = \frac{TP}{TP + FN}
\label{eq:3}
\end{equation}

The F1-score is the harmonic mean of precision and recall, providing a balance between the two. It is particularly useful when the class distribution is imbalanced.

\begin{equation}
\text{F1-Score} = 2 \times \frac{\text{Precision} \times \text{Recall}}{\text{Precision} + \text{Recall}}
\label{eq:3}
\end{equation}

These metrics together provide a comprehensive evaluation of the classification performance, helping us to identify the most effective method.

To fully characterize our \texttt{CatBoostClassifier} behavior across all decision thresholds, we analyze the precision-recall curve \citep[e.g., ][]{cook2020consult}. As shown in Figure~\ref{fig:fig4_roc}, the high area under the curve (AUC = 0.97) confirms robust performance on our imbalanced dataset. At our chosen threshold of 0.5, the model achieves 90\% precision while maintaining 88\% recall - an optimal balance for identifying quiescent galaxies without excessive contamination. This balance is critical because our primary goal is to accurately identify as many quiescent galaxies as possible (high recall) while minimizing the number of star-forming galaxies that are incorrectly classified as quiescent (high precision).

\begin{table*}
\centering
\begin{tabular}{c|c|ccc|c}
\hline \hline
\toprule

Method & Class & Precision & Recall & F1-Score & Execution time\\
\midrule
Machine Learning & Quiescent &$0.899$&$0.877$&$0.888$& $\sim{10s}$\\
                 & Star-Forming &$0.980$&$0.984$&$0.982$&\\
\midrule
SED-fitting      & Quiescent &$0.953$&$0.375$&$0.538$& $\sim{25,000s}$\\
                 & Star-Forming &$0.909$&$0.997$&$0.951$&\\
\bottomrule
\end{tabular}
\caption{Comparison of the performance of ML and SED-fitting classifiers on the testing set of star-forming and quiescent galaxies, evaluated using precision (purity), recall (completeness), and F1-score metrics.}
\label{tab:tab2}
\end{table*}

\begin{figure*}
    \centering
    \includegraphics[width=\linewidth]{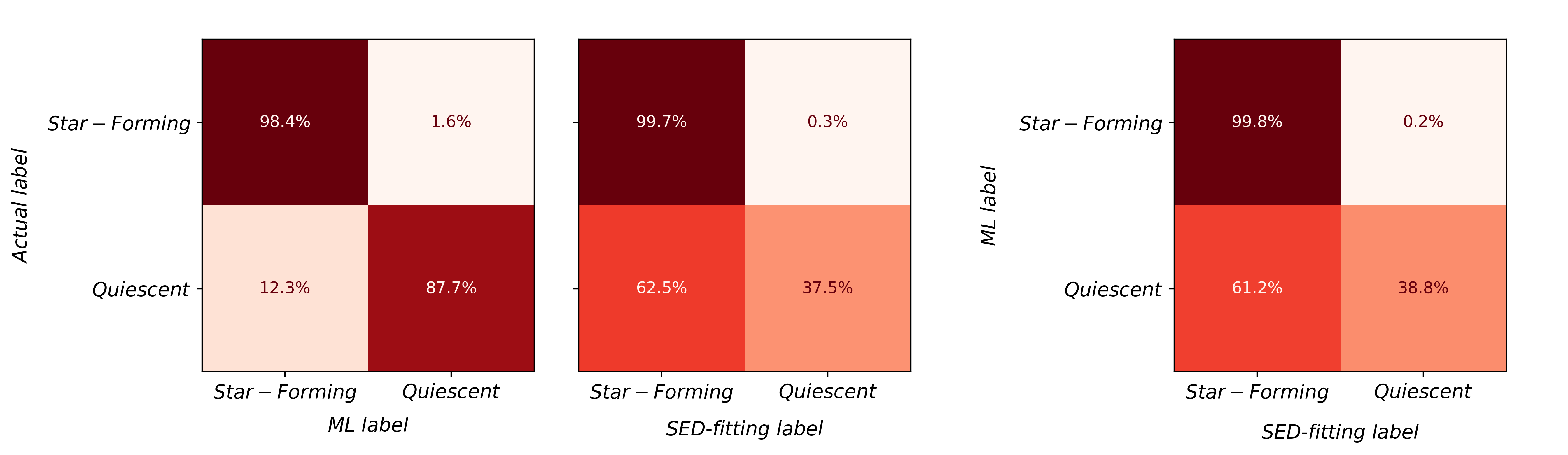}
    \caption{Confusion matrices compare the predicted classifications from different methods against the ground truth and each other. The matrices display: (left) ML predictions versus the actual labels, (middle) SED-fitting predictions versus the actual labels, and (right) a direct comparison of the predictions from the ML and SED-fitting methods. The actual labels are derived from the mock catalog described in Section~\ref{sub:sub1}.}
    \label{fig:fig4}
\end{figure*}

\begin{figure}
    \centering
    \includegraphics[width=0.88\linewidth]{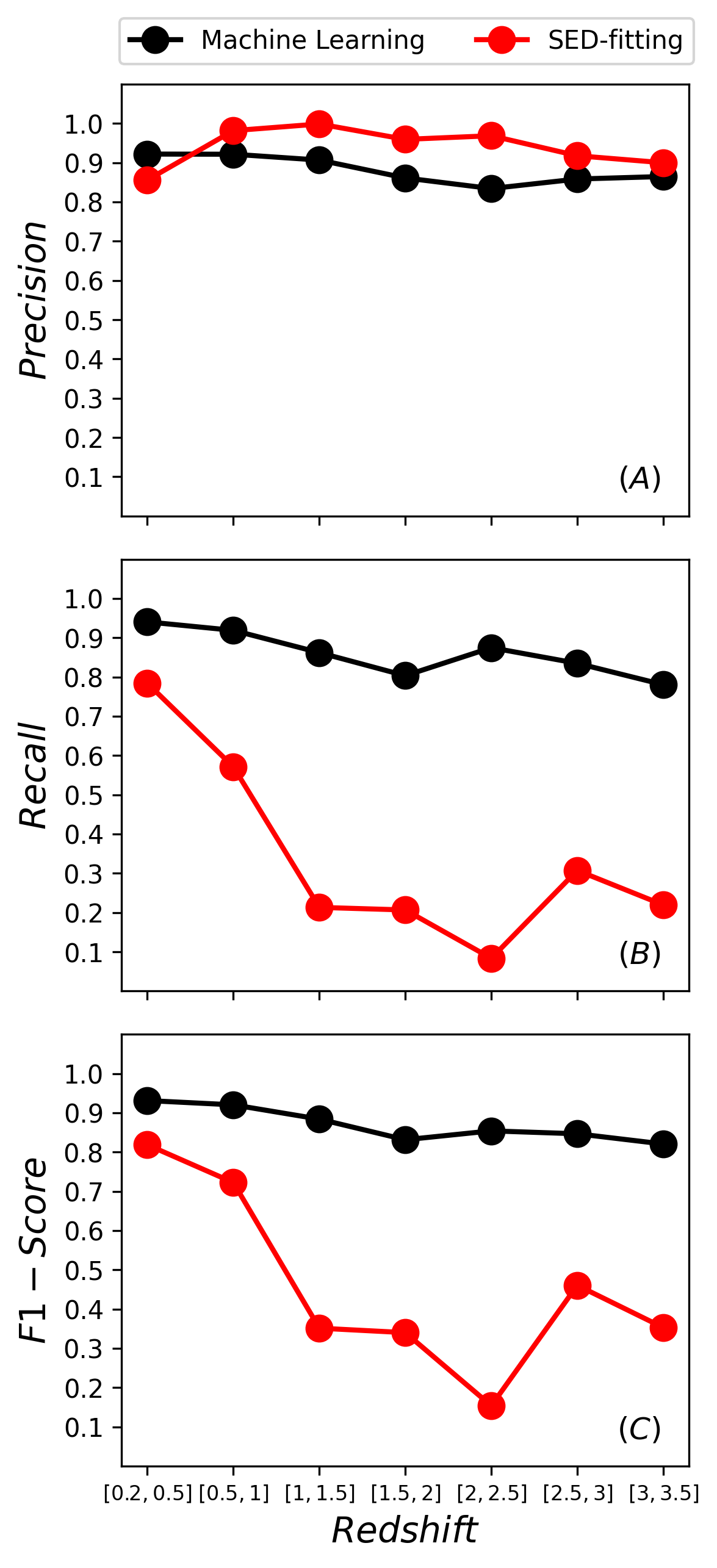}
    \caption{precision (purity), recall (completeness), and F1-score of ML and SED-fitting classifiers on the testing set for quiescent galaxies as a function of redshift bins.}
    \label{fig:fig5}
\end{figure}

\subsection{Comparison of classification methods}\label{sub:sub2}
Table~\ref{tab:tab2} presents the performance of both ML and SED-fitting classifiers on the testing set. We evaluated the classifiers using standard metrics: precision, recall, and F1-score. As expected, both classifiers achieve high performance for star-forming galaxies, with F1-scores of 98.2\%(ML) and 95.1\%(SED-fitting). This is likely due to the significantly larger sample size of star-forming galaxies (20,968) compared to quiescent galaxies (2,111).

However, for quiescent galaxies, the ML method significantly outperforms the SED-fitting approach, achieving an F1-score of 88.8\% compared to 53.8\%. A closer examination reveals that both methods exhibit high precision, accurately identifying most predicted quiescent galaxies. In fact, the SED-fitting method achieves slightly higher precision than the ML method, with a difference of approximately 5 percentage points.

Nevertheless, the ML method substantially surpasses the SED-fitting approach in recall, correctly identifying 87.7\% of actual quiescent galaxies compared to 37.5\%. This 50-percentage-point difference indicates that the SED-fitting method misses a substantial number of quiescent galaxies. Figure~\ref{fig:fig4} presents the confusion matrices for both methods, further illustrating these differences. The SED-fitting method misclassifies 62.5\% of quiescent galaxies as star-forming, whereas the ML method misclassifies only 12.3\%.

Additionally, the ML method offers a significant speed advantage, executing over 2,500 times faster than the SED-fitting method (10s vs. 25,000s) on an 11th Gen Intel® Core™ i7-1165G7 processor.

To delve deeper into the performance of these two methods, we selected quiescent galaxies in redshift bins delimited by the values z = 0.2, 0.5, 1, 1.5, 2, 2.5, 3 and 3.5. We then calculated precision, recall, and F1-score for different redshift bins for quiescent galaxies using both methods. Figure~\ref{fig:fig5} displays these metrics as a function of redshift. While the performance of the ML method remains relatively constant, the SED-fitting method exhibits significant fluctuations (see panel (C)).

Specifically, for precision (panel A), the SED-fitting method generally outperforms the ML method, except within the 0.2-0.5 redshift range. This indicates that the purity of the SED-fitting quiescent galaxy sample is generally higher. For recall (panel B), the ML method outperforms the SED-fitting method across all redshift ranges, particularly at higher redshifts. In fact, the completeness of the quiescent galaxy sample predicted by the SED-fitting method, except within the 0-1 redshift range, is dramatically lower than that of the ML method. For instance, within the 2-2.5 redshift range, the difference in recall reaches approximately 80 percentage points. Table~\ref{tab:tab3} summarizes the metric performance of both methods for the classification of quiescent galaxies in different redshift bins.


\section{Discussion}\label{sec:6}
In this study, we employed both ML and spectral energy distribution (SED)-fitting techniques to classify galaxies as either quiescent or star-forming. We utilized simulations from the Santa Cruz semi-analytical modeling framework for this purpose. Our goal was to compare the performance of these methods using metrics such as precision (purity), recall (completeness), and the F1-score, with the ultimate aim of applying the superior method to the \texttt{Farmer} COSMOS2020 galaxy catalog.

To bridge the gap between the simulated and observational data, we implemented several pre-processing steps. First, we identified eight mutual bands that were present in both catalogs. For the COSMOS2020 data, we utilized the galaxy sample constructed by \cite{weaver2023cosmos2020}, which was drawn from the contiguous 1.27 deg$^2$ COMBINED region. This sample underwent stringent quality cuts, including the removal of sources with low infrared detection ($m_{\text{ch1}} > 26$ AB), ambiguous redshifts, and unreliable SED fits. For the mock catalog, we focused on the COSMOS field's zeroth realization, applying a similar magnitude cut ($m_{\text{ch1}} < 26$ AB).

Our analysis focused on galaxies within the redshift range $0.2 < z < 3.5$ for both datasets. This restriction was necessary to avoid the scarcity of quiescent galaxies at $z > 3.5$, which would have negatively impacted our ML method for quiescent galaxy classification. To ensure the completeness of our galaxy samples, we adopted the mass completeness limits for star-forming and quiescent galaxies derived by \cite{weaver2023cosmos2020} (see Equations~\ref{eq0} and \ref{eq00}).

To address missing values in the COSMOS2020 sample, we used \texttt{MissForest} method.  To further enhance the realism of the simulated data, we injected noise by simulating observational errors based on the uncertainties observed in the COSMOS2020 catalog. Finally, we identified quiescent galaxies in the simulated catalog using an evolving $sSFR$ threshold (see Equation~\ref{eq1}).

\begin{table}
\centering
\begin{tabular}{c|c|ccc}
\hline \hline
\toprule

Method & Redshift Range & Precision & Recall & F1-Score \\
\midrule
Machine  & 0.2-0.5 &$0.922$&$0.940$&$0.931$ \\
              Learning   & 0.5-1 &$0.922$&$0.919$&$0.920$ \\
                 & 1-1.5 &$0.907$&$0.862$&$0.884$ \\
                 & 1.5-2 &$0.861$&$0.804$&$0.831$ \\
                 & 2-2.5 &$0.834$&$0.875$&$0.854$ \\
                 & 2.5-3 &$0.859$&$0.835$&$0.847$ \\
                 & 3-3.5 &$0.865$&$0.781$&$0.821$ \\

\midrule
SED-fitting      & 0.2-0.5 &$0.856$&$0.784$&$0.819$ \\
                 & 0.5-1 &$0.982$&$0.571$&$0.722$ \\
                 & 1-1.5 &$0.999$&$0.213$&$0.351$ \\
                 & 1.5-2 &$0.960$&$0.207$&$0.340$ \\
                 & 2-2.5 &$0.969$&$0.084$&$0.154$ \\
                 & 2.5-3 &$0.918$&$0.307$&$0.461$ \\
                 & 3-3.5 &$0.900$&$0.220$&$0.353$ \\

\bottomrule
\end{tabular}
\caption{Performance metrics of ML and SED-fitting classifiers for quiescent galaxies in different redshift bins, including precision (purity), recall (completeness), and F1-score.}
\label{tab:tab3}
\end{table}

The resulting simulated galaxy catalog was divided into training and testing sets. A \texttt{CatBoostClassifier} was trained on the training set, and its performance was compared to that of the SED-fitting approach on the testing set.

Our study demonstrates that the ML method significantly outperforms the SED-fitting method in classifying quiescent galaxies in both accuracy and execution time.

Regarding execution time, SED-fitting is significantly slower than ML classifier algorithms such as \texttt{CatBoostClassifier}. SED-fitting analyzes each galaxy in testing set independently, comparing its spectrum to a vast library of approximately 7 million model galaxies through an iterative fitting process. This individual analysis and extensive comparison make it computationally intensive, especially for large datasets. In contrast, \texttt{CatBoostClassifier} processes the entire testing set in batches, leveraging a pre-trained model to make efficient predictions. By avoiding individual galaxy analysis and iterative fitting, \texttt{CatBoostClassifier} offers a significantly faster approach, exceeding the speed of the SED-fitting method by more than three orders of magnitude on CPU implementations, making the ML method considerably more suitable for large-scale applications where computational speed is crucial \cite[see for more details ][]{asadi2025leveraging}.

It is important to mention that we excluded pre-processing steps from our timing comparison because these steps are highly variable and application-dependent. For example, in SED-fitting method, the time required for template construction can vary significantly based on the chosen parameters, such as the inclusion of additional dust laws, different star formation histories, or finer redshift steps. Similarly, in ML method, pre-processing time can change considerably depending on the hyperparameters and their ranges. Some studies may not require imputation, noise injection, or larger training set sizes, further highlighting this variability. By focusing on the inference computational tasks, our analysis provides a more consistent and focused comparison.

The superiority of the ML method in terms of accuracy can be attributed to several factors:
\begin{itemize}
    \item The SED-fitting model often uses simplified models for star-formation histories when fitting galaxies. This simplification contrasts with the more complex star-formation histories present in the SAM catalog. As a result, the SED-fitting classifier may not accurately capture the nuances of galaxy evolution. This issue is particularly pronounced for galaxies that have recently become quiescent or are in the final stages of this transition. These galaxies' SEDs are more sensitive to recent star-formation episodes, where type A stars still significantly contribute to the total SED. Moreover, young dusty galaxies can exhibit colors similar to those of old, dust-free quiescent galaxies, making it challenging to distinguish between the two. While the accuracy of the SED-fitting method can be improved using more sophisticated codes, such as Bayesian SED-fitting codes \citep[e.g.,] []{johnson2021stellar, hahn2022accelerated} or those use more realistic star formation histories \citep[e.g., see] []{leja2019measure}, this often comes at the cost of increased execution time as the number of model sources rises significantly. 
    
    \item By sidestepping the need to explicitly derive physical properties like sSFR, the ML method offers a more streamlined approach to galaxy classification. Instead of relying on complex models to translate observed light into physical parameters, it directly learns the relationship between a galaxy's light and its type. This avoids potential inaccuracies that can arise from the simplifying assumptions inherent in SED-fitting models, leading to a more robust and efficient classification process.
    
    \item Unlike SED-fitting, which operates on individual galaxy SEDs, our ML model learns the discriminatory power of each color across the entire training population. Colors with low discriminatory power for the classification task are effectively down-weighted during training.

    \item While the SED-fitting method prioritizes accurate physical parameter estimation, the ML model is specifically designed to optimize classification decisions. This focus on overall classification accuracy can lead to superior performance, even if individual physical parameter estimates may be less precise.
\end{itemize}

The effectiveness of our ML model depends on the quality and representativeness of the semi-analytic training galaxies relative to COSMOS2020 galaxies. While promising, the true robustness of our approach will be enhanced by using confirmed representative samples for training. Focusing on a carefully verified subset of observations will improve the accuracy and reliability of our model when applied to large-scale surveys. Furthermore, validating the model against diverse simulation catalogs will strengthen our findings and provide a robust framework for future research.

Additionally, in this study, we had to restrict our analysis to 8 mutual bands found between the mock and COSMOS2020 catalogs. Having more mutual bands would have resulted in a more robust ML model and better predictions. Therefore, using simulation data that are more consistent with observational data is crucial.

\begin{figure}
    \centering
    \includegraphics[width=\linewidth]{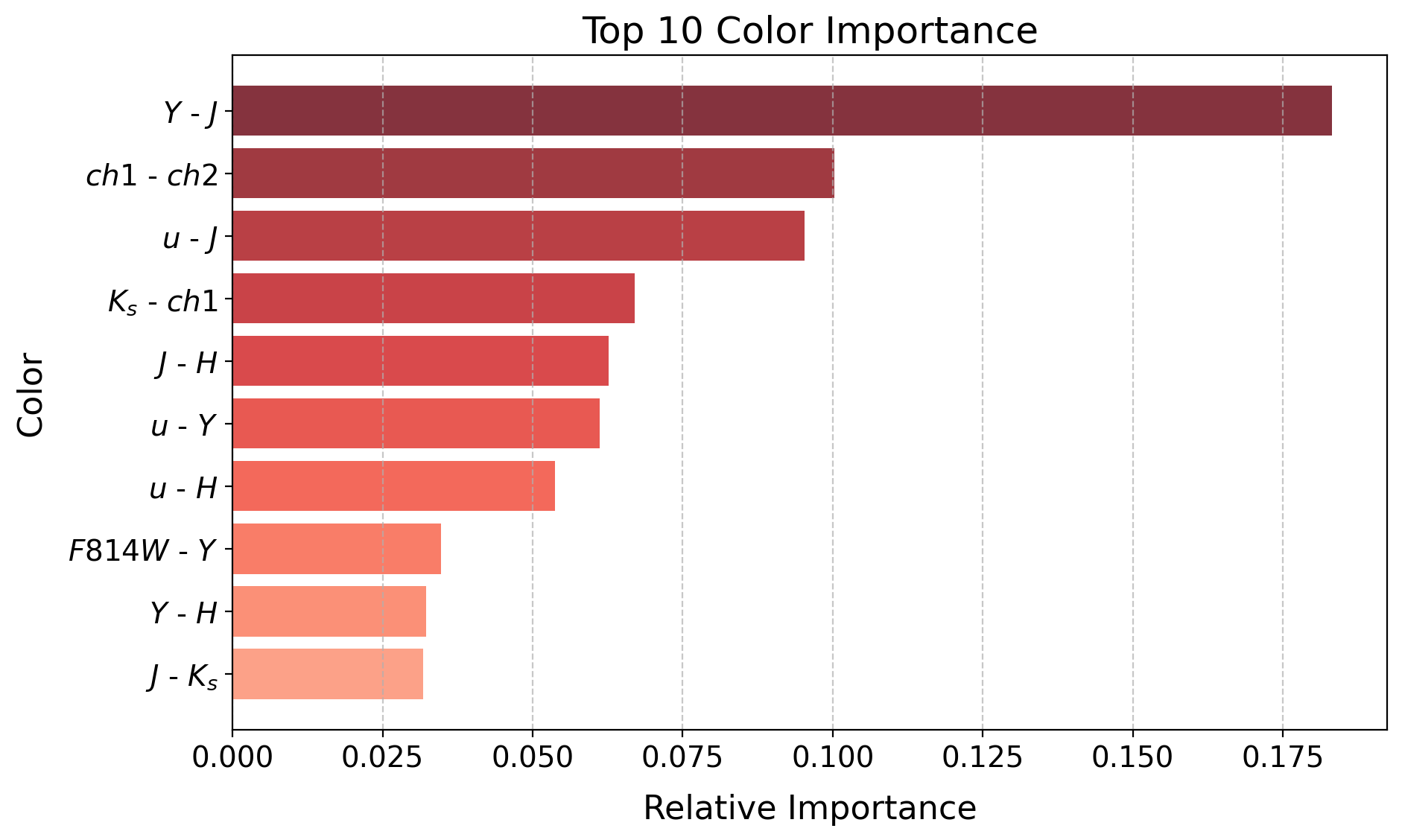}
    \caption{Relative importance of the top ten colors (of 28 total color features) in the trained model for classifying quiescent and star-forming galaxies in the COSMOS2020 sample.}
    \label{fig:fig6}
\end{figure}


\section{Classification of COSMOS2020 galaxies}\label{sec:7}
To apply our ML method to the COSMOS2020 sample, which consists of 365,877 galaxies, we first trained a \texttt{CatBoostClassifier} on the complete mock sample, which included all eight realizations of the data. The training utilized 28 color features derived from the selected bands (Table~\ref{tab:tab0}). Prior to training, we standardized the features using the \texttt{StandardScaler} function to ensure consistent scaling across the dataset. Following this, we applied the same approach to the COSMOS2020 dataset. Specifically, we constructed analogous color features, standardized them in the same manner, and used the trained model to classify galaxies as either quiescent or star-forming.

\subsection{Color importance}
Before evaluating the classification result of the trained model, we first examined the importance of the color features used by the model. To do this, we employed the split count method provided by the \texttt{CatBoostClassifier}. This method calculates color importance by counting how often each color is used to split the mock data across all the decision trees in the ensemble. Colors that are used more frequently for splitting are considered more important because they play a larger role in dividing the data and making predictions.

Figure~\ref{fig:fig6} illustrates the relative importance of the top ten color features in our trained model. As shown, the Y - J color feature is the most important, followed by ch1 - ch2 and u - J. These features are likely significant because they capture key spectral properties that distinguish quiescent galaxies from star-forming ones. For example, the Y - J color is sensitive to the Balmer break, a hallmark of quiescent galaxies, as it indicates the presence of older stellar populations with minimal ongoing star formation. On the other hand, the ch1 - ch2 color, derived from mid-infrared bands, is sensitive to dust emission, which is typically more prominent in star-forming galaxies due to their higher rates of dust production and reprocessed light from young, hot stars.

\begin{figure}
    \centering
    \includegraphics[width=0.93\linewidth]{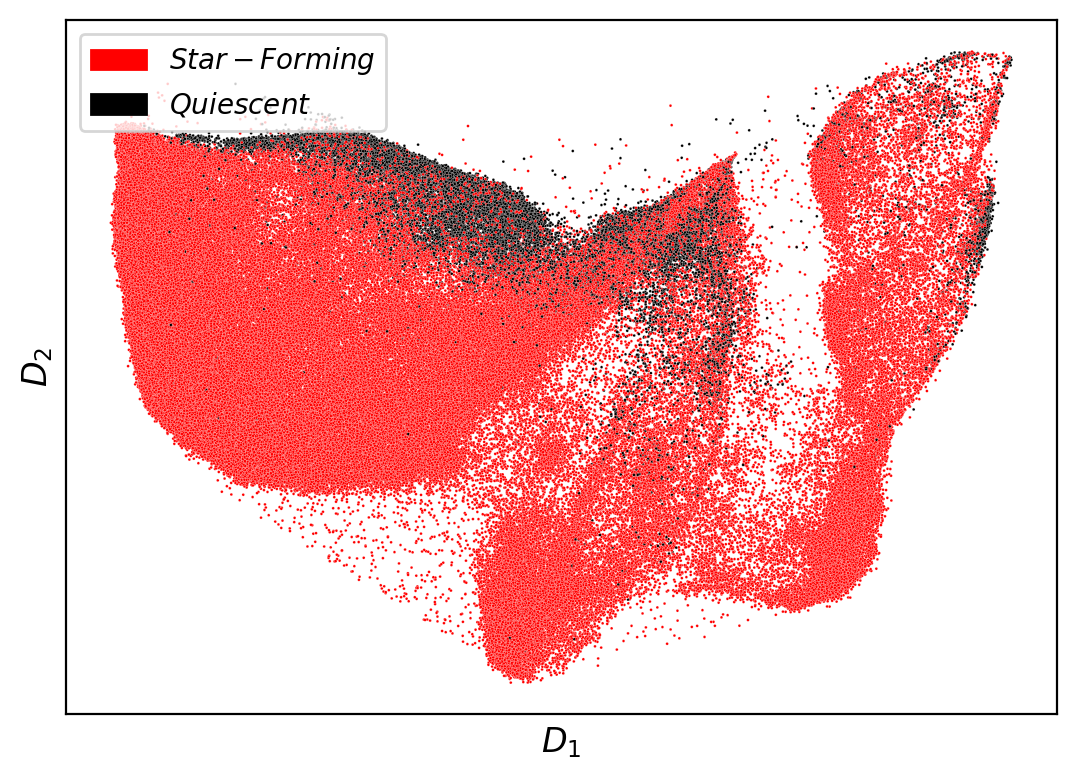}
    \caption{A 2D UMAP projection provides a clear visualization of the distinct populations of quiescent and star-forming galaxies using ML within the COSMOS2020 sample data. The dimensions of D1 and D2 are arbitrary labels and do not carry any physical significance.}
    \label{fig:fig7}
\end{figure}

\subsection{Classification result}
A 2D UMAP projection of the COSMOS2020 sample, shown in Figure~\ref{fig:fig7}, visually distinguishes between quiescent and star-forming galaxies based on our trained model. To evaluate the performance of our ML approach, we compared it with two widely used methods: (1) an SED-fitting method using \texttt{LePhare} median sSFR and Equation~\ref{eq1}, and (2) the NUVrJ method, which has been applied to this sample by \cite{weaver2023cosmos2020}.

\begin{table}
\centering
\begin{tabular}{c|c|ccc}
\hline \hline
\toprule

Method           & Class        & Precision & Recall  & F1-Score \\
\midrule
SED-fitting      & Quiescent    & $0.906$   & $0.508$ & $0.651$ \\
                 & Star-Forming & $0.929$   & $0.992$ & $0.960$  \\
\midrule
NUVrJ            & Quiescent    & $0.923$   & $0.462$ & $0.616$  \\
                 & Star-Forming & $0.923$   & $0.994$ & $0.957$   \\
\bottomrule
\end{tabular}
\caption{Precision (purity), recall (completeness), and F1-score for quiescent and star-forming galaxies, comparing the SED-fitting and NUVrJ \citep{weaver2023cosmos2020} methods to the ML method. High precision values ($>90\%$) indicate strong compatibility in quiescent galaxy predictions, while lower recall values ($\sim$50\%) suggest that both methods miss a significant fraction of quiescent galaxies identified by the ML approach.}
\label{tab:tab4}
\end{table}

Table~\ref{tab:tab4} summarizes the precision (purity), recall (completeness), and F1-score for quiescent and star-forming galaxies, comparing the SED-fitting and NUVrJ methods to the ML method. For star-forming galaxies, both methods show strong agreement with the ML method, achieving F1-scores greater than 95\%. For quiescent galaxies, both methods exhibit high precision, exceeding 90\%, indicating that the majority of their quiescent predictions align with those of the ML method. However, the recall for quiescent galaxies is approximately 50\%, suggesting that only half of the quiescent galaxies identified by the ML method were detected by the other methods. This implies that while the SED-fitting and NUVrJ methods are highly compatible in their quiescent predictions (high precision), they miss a significant portion of the quiescent galaxies identified by the ML approach (low recall).

\begin{figure}
    \centering
    \includegraphics[width=\linewidth]{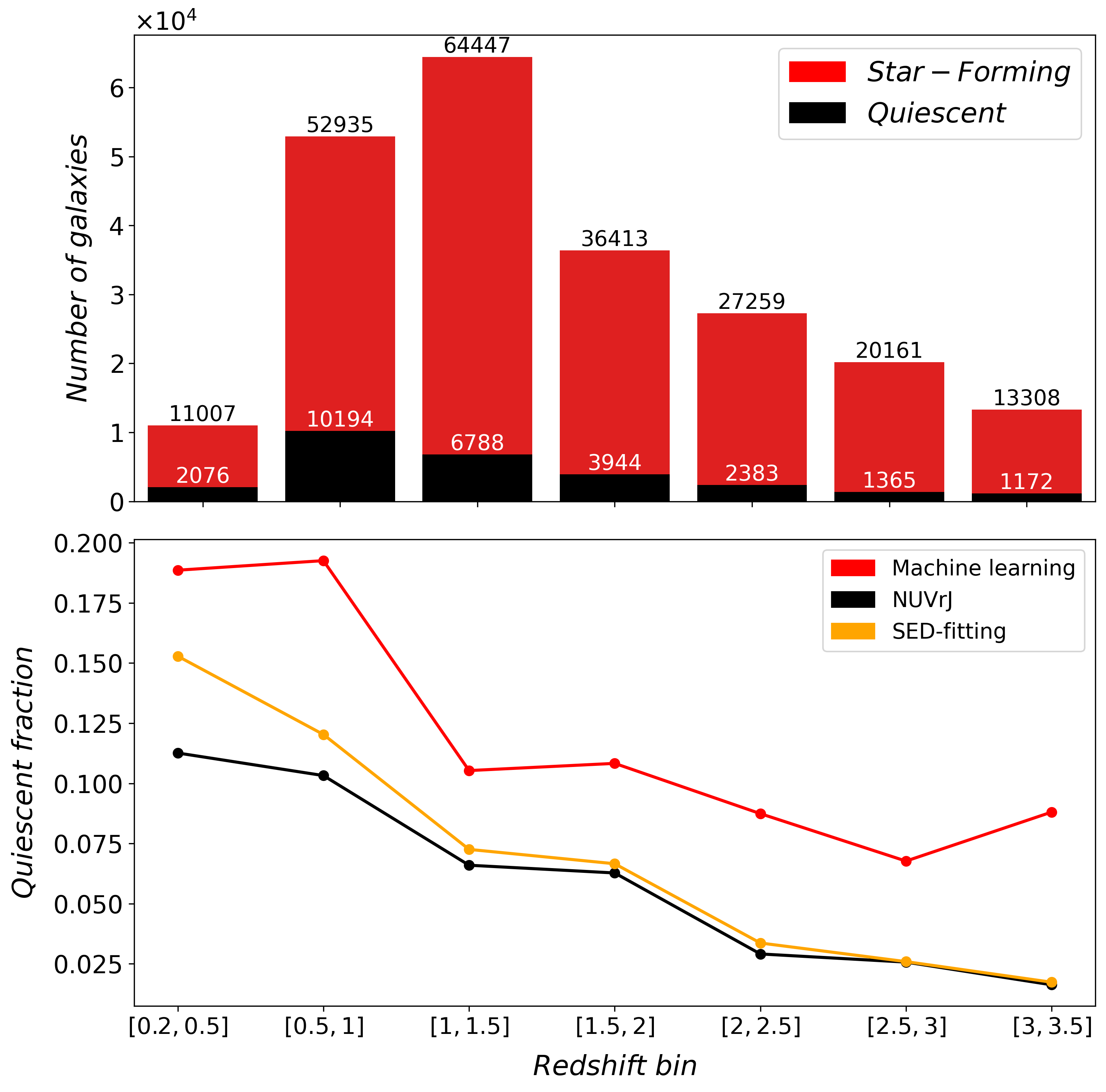}
    \caption{Upper panel: Distribution of quiescent galaxies and the total number of galaxies across various redshift bins, as determined by the ML method. Lower panel: Evolution of the fraction of quiescent galaxies as a function of redshift for galaxies with stellar masses $M_{QG} \geq 10^{9.5} M_{\odot}$, comparing the NUVrJ \citep{weaver2023cosmos2020}, SED-fitting, and ML methods on the COSMOS2020 data sample.}
    \label{fig:fig8}
\end{figure}

In Figure~\ref{fig:fig8}, the upper panel illustrates the distribution of quiescent galaxies and the total number of galaxies across various redshift bins, as determined by the ML method. The lower panel presents the evolution of the fraction of quiescent galaxies as a function of redshift for galaxies with stellar masses $M_{QG} \geq 10^{9.5} M_{\odot}$, comparing the NUVrJ, SED-fitting, and ML methods on the COSMOS2020 dataset. As shown, the NUVrJ and SED-fitting methods exhibit similar trends across the redshift bins. In contrast, the ML method predicts a systematically higher fraction of quiescent galaxies in all redshift bins compared to the other two methods.

This result is consistent with what we obtained in Section~\ref{sub:sub2}, where we compared the performance of the ML method with the SED-fitting method on the mock data sample. In the simulation, the ML method demonstrated superior recall for quiescent galaxies compared to the SED-fitting approach, highlighting its ability to identify true quiescent galaxies more effectively. This aligns with its performance on the real COSMOS2020 data, where it also achieves higher recall for quiescent galaxies than traditional methods.

\section{Conclusion}\label{sec:8}
In this study, we developed and validated a ML framework for classifying quiescent and star-forming galaxies within the redshift range $0.2<z<3.5$, leveraging mock photometric data from the Santa Cruz semi-analytic model (SAM) and applying it to the \texttt{Farmer} COSMOS2020 catalog. Our approach utilized a \texttt{CatBoostClassifier} trained on 28 color features derived from eight mutual bands, enabling efficient and accurate classification. By comparing this method to traditional SED-fitting approach in the simulation space, we demonstrated its superiority in identifying quiescent galaxies, particularly at high redshifts ($z>1$), where the SED-fitting method exhibits significant incompleteness. Key findings include:
  \begin{itemize}
     \item Performance superiority: The ML classifier achieved an F1-score of 89\% for quiescent galaxies, significantly outperforming the SED-fitting method (54\%). This improvement is primarily due to a notable difference in recall (completeness), with the ML method identifying 88\% of quiescent galaxies compared to just 38\% for the SED-fitting method. While precision (purity) rates were comparable, the SED-fitting method performed slightly better in this regard.

    \item Computational efficiency: The ML approach executed classifications over 2,500 times faster than SED-fitting, making it practical for large-scale surveys.

    \item Application to observations: When applied to the COSMOS2020 catalog, the ML model identified a systematically higher fraction of quiescent galaxies across all redshift bins within $0.2<z<3.5$ compared to both the NUVrJ and SED-fitting methods. This result is consistent with the simulation findings, where the ML model demonstrated superior completeness, particularly for high-redshift quiescent populations.
\end{itemize}

The methodology developed in this study for classifying quiescent galaxies is adaptable and can be extended to various other galaxy types. This flexibility enables the exploration of diverse galaxy populations, such as Lyman Break galaxies or active galactic nuclei. By tailoring our ML method to these different groups, researchers can delve deeper into the unique physical processes and evolutionary pathways that characterize each population, ultimately contributing to a more comprehensive understanding of galaxy formation and evolution in the universe.

Looking ahead, the integration of ML techniques, such as the one demonstrated in this study, holds immense potential for revolutionizing galaxy classification. As datasets continue to grow in size and complexity with upcoming surveys, the scalability, accuracy, and efficiency of ML methods will become increasingly indispensable. By bridging the gap between simulations and observations, these approaches not only enhance our ability to classify galaxies but also open new avenues for exploring the underlying physics of galaxy evolution.

\section*{Data Availability}\label{sec:data}
The trained ML model and full classified COSMOS2020 catalog (including quiescent and star-forming galaxy classifications) resulting from this analysis are publicly available in the GitHub repository: \url{https://github.com/vahidoo7/ML-Quiescent-Galaxy-Classifier}.

\software{Pandas \citep{mckinney2011pandas}, Scikit-learn \citep{pedregosa2011scikit}, Numpy \citep{harris2020array}, UMAP \citep{mcinnes2018umap}, Catboost \citep{prokhorenkova2018catboost}, Astropy \citep{robitaille2013astropy}, Seaborn \citep{waskom2021seaborn}, Matplotlib \citep{barrett2005matplotlib}, SciPy \citep{virtanen2020scipy}, MissForest \citep{stekhoven2015missforest}, LePhare \citep{arnouts1999measuring}.}

\bibliography{ref}

\end{document}